\documentclass[a4paper]{article}
\usepackage[T1]{fontenc}
\usepackage[utf8]{inputenc}
\DeclareUnicodeCharacter{0301}{\'{e}}
\usepackage[english]{babel}
\usepackage{graphicx}
\usepackage{hyperref}
\hypersetup{
  setpagesize  = false,
  colorlinks   = true,    
  urlcolor     = blue,    
  linkcolor    = black,   
  citecolor    = black    
}
\usepackage{authblk}
\usepackage{geometry}
\usepackage{setspace}
\usepackage{verbatim}
\usepackage{mathtools}
\usepackage[utf8]{inputenc}
\usepackage{amsmath}
\usepackage{cleveref}

\usepackage{amssymb}
\usepackage{gensymb}
\usepackage{verbatim} 
\usepackage{ragged2e} 
\usepackage[english]{babel}
\usepackage{amsfonts}
\usepackage{cancel}

\usepackage[backend=biber]{biblatex}
\newcommand{\keywordsenglishname}{Keywords}

\renewenvironment{abstract}{%
        \begin{center}
	\begin{minipage}{14cm}
	{\textbf{\abstractname:}}
}{
        \end{minipage}
	\end{center}
}

\title {Liquidity Provision Payoff on Automated Market Makers\\[1ex] }
\author{Jin Hong Kuan \thanks{E-mail address: jin@eastrocklabs.com} $^1$ }
\affil{East Rock Labs}
\date{}

\usepackage{fancyhdr}
\begin{document}

\maketitle
\vspace{6pt}




\section{Abstract}

The standard approach for compensating liquidity providers on many decentralized exchanges (DEX) for serving as counter-party to swaps is through charging a small percentage of fees. The expected payoff from the cash flow of this mode of market making has yet to be mathematically formulated in terms of volatility in the existing literature. We provide here a preliminary derivation of the payoff formula, by making the standard set of assumptions for efficient markets, namely geometric Brownian price movements and zero arbitrage. Trading volume, conventionally taken as an exogenous variable for fees calculation, becomes a function of volatility and available liquidity in this formulation.  In doing so, we show that it is a near-linear function of the volatility of the underlying risky asset. Since hedging instruments with such a property is highly sought after, we discuss the potential of securitizing the cash flow of liquidity fees to serve as a volatility product in its own right.  

\section{Derivations} 

We model here the relationship between the cash flow from liquidity fees with the relative price movements between two tokens $(a,b)$ in a constant product market maker (CPMM). To standardize calculations, all transactions are notarized in units of $a$, and we assume that fees will always be paid in that token. Moreover, all the liquidity is assumed to be unchanged in all calculations below, i.e. no liquidity provision takes place. Let the reserve of each token be denoted by $r_a, r_b$ respectively. The initial liquidity $L^2$, a quantity that remains invariant after each swap is then defined as follows:
\begin{align*}
    L^2 &= r_{a,0} \cdot r_{b,0} 
\end{align*}
The relative price in a CPMM is fixed as the ratio between the respective reserves of each token:
\begin{align}
    P_t &= \frac{r_{b,t}}{r_{a,t}} \nonumber \\
    &= \frac{L^2}{r^{2}_{a,t}} \nonumber 
\end{align}
Since a percentage fee $\gamma$ is charged for each transaction and deposited into the fees vault, the change in fees collected $dF$ is proportional to the inflow and outflow of token $a$. While each market makers may differ slightly in implementation, we capture this commission-based revenue model with a simple model:
\begin{align}
    dF &= \gamma \lvert dr_a \rvert \label{eq:dF} 
\end{align}
The absolute operator is included since a positive amount of fee is collected regardless of the direction of the swap. 

\paragraph{Instantaneous Payoff Rate.} We now attempt to derive $dr_a$, which we know to be dependent on price. Specifically, manipulating the algebraic relationship defined above between price and token reserve, we obtain
\begin{align}
     r_{a,t} &=LP_t^{-1/2}\label{eq:r_aP}
\end{align}
Suppose that the relative price follows a geometric Brownian process, commonly assumed for asset price movements in the extensive continuous-time financial modeling literature \cite{ross_continuous_1991}: 
\begin{align}
    dP_t &= \mu P_t dt + \sigma P_t dz \label{eq:geom}
\end{align}
where $dz = \epsilon \sqrt{dt} $ is the stochastic term, and $\epsilon \sim \mathcal{N}(0,1)$ is a standard Wiener process. Note that the volatility term can be interpreted as the gross volatility resulting from reaction to market events as well as noise trades. We make no distinction between informed and uninformed trades.

In a well-arbitraged market, future price drift is priced in, thus we can assume $P_t$ to be a martingale, and $\mu = 0$. Combining \cref{eq:geom} with \cref{eq:r_aP} in accordance to Itô's lemma allows us to then derive the stochastic process governing $dr_a$: 
\begin{align}
    dr_a &= \left( \frac{\partial r_a}{\partial t} + \mu\frac{\partial r_a}{\partial P} + \frac{\sigma^2}{2} \frac{\partial^2 r_a}{\partial P^2} \right) dt + \sigma \frac{\partial r_a}{\partial P} dz \nonumber \\ 
    &= \left( 0 -  \frac{\mu}{2} L P^{-3/2} + \frac{3\sigma^2}{4} LP^{-5/2} \right) dt -  \frac{\sigma}{2} L P^{-3/2} dz \nonumber \\ 
    &=  L P^{-3/2} \left(  \frac{3\sigma^2 }{4P}  dt -  \frac{\sigma}{2} dz \right)
\end{align}
And consequently, the expected payoff rate per unit of liquidity (i.e. divided by total $L$), which we henceforth will refer to as payoff rate, is a price-dependent stochastic differential: 
\begin{align}
    dF(P)/L &= \gamma \lvert dr_a(P) \rvert / L \nonumber \\ &=\gamma P^{-3/2} \bigg\lvert  \frac{3\sigma^2}{4P}  dt -  \frac{\sigma}{2} dz \bigg\rvert 
\end{align}
Given a particular $P$, the payoff rate is the first moment of the above equation:
\begin{align*}
    E[d F(P)/L]  &=\gamma P^{-3/2} \int_{-\infty}^{\infty}  \bigg\lvert  \frac{3\sigma^2 }{4P}  dt -  \frac{\sigma \epsilon}{2} \sqrt{ dt}  \bigg\rvert \phi (\epsilon) d\epsilon 
\end{align*}
where $\phi(\cdot)$ is the probability density function of a unit normal distribution. 

We solve the integration by comparison. For any constant $A, B$, it can be shown that
\begin{align*}
    \int_{-\infty}^{\infty} \lvert A-B\epsilon \rvert \phi(\epsilon) d\epsilon = \sqrt{\pi} \left[ 2A\phi \left(\frac{B}{A} \right) + 2B\Phi \left(\frac{B}{A} \right) - B \right]
\end{align*}
where $\Phi$ is the cumulative density function of a unit normal distribution. Hence, we have
\begin{align}
    E[dF(P)/L] &=  \frac{\gamma}{\sqrt{\pi P^3}} \left[ 2A\phi \left( \frac{B}{A} \right) + 2B\Phi \left(\frac{B}{A} \right) - B \right] \label{eq:payoff}\\
    A &=  \frac{3dt}{4P}    \sigma^2 \nonumber\\ 
    B &=  \frac{\sqrt{ dt}}{2}   \sigma \nonumber
\end{align}
One notable property about the expected payoff function is that it can be shown to be \emph{almost linear} to volatility $\sigma$ within the reasonable range of the parameter. This hints at the potential utility of liquidity swaps as direct volatility hedge or bet.  

To show this, we first take the derivatives of each term with respect to $\sigma$.  We separate out the constant coefficients, such that $A = \alpha \sigma^2$, $B = \beta \sigma$ and $\lambda = \beta/\alpha$: 
\begin{equation*}
\begin{aligned}
     \partial_\sigma A  &= 2\alpha \sigma \\ 
      \partial_\sigma B &= \beta
\end{aligned}
\qquad 
\begin{aligned}
    \left. \partial_\sigma \Phi \right|_{\frac{\lambda}{\sigma}} &= - \lambda \phi \left( \frac{\lambda}{\sigma} \right) \\
    \left. \partial_\sigma \phi \right|_{\frac{\lambda}{\sigma}} &= \lambda \sigma \phi \left( \frac{\lambda}{ \sigma} \right) 
\end{aligned}
\end{equation*}
Applying a series of chain rules on \cref{eq:payoff}, we obtain
\begin{align}
    \partial_\sigma E[dF(P)/L]  = \sigma \alpha  \phi \left(\frac{\lambda}{\sigma} \right) +  \beta \Phi(\frac{\lambda}{\sigma})  
\end{align}
This measure is commonly known as vega, which stands for the sensitivity of returns to volatility. For $\frac{\lambda}{\sigma} \gg 1, \phi(x) \approx 0, \Phi(x) \approx 1$, the second term dominates. Thus, $\partial E / \partial \sigma \approx \beta $ and the payoff rate of liquidity is almost linear to volatility. Moreover, since $\beta$ does not depend on price, this implies that payoff rate should remain relatively constant at all prices when volatility is low. On the contrary, in high volatility regimes where $\frac{\lambda}{\sigma}$ is close to zero, the first term becomes increasingly prominent and  $\partial E / \partial \sigma \approx \alpha \sigma$. In this case, the payoff rate of liquidity swaps becomes \emph{quadratic} to volatility. 

It is imperative to point out that $\lambda \rightarrow \infty$ as $dt \rightarrow 0$, which in theory renders any consideration of the $\sigma \alpha$ term moot. An intuitive explanation of this phenomenon is that since the stochastic term dominates the drift term in infinitesimally short time intervals, a sign switch is applied by the absolute value operator about half the time, causing the drift terms to cancel each other out when summed. However this assumes a perfectly smooth stochastic process and near-instant arbitrage, which does not realistically represent the market. Hence, we keep all relevant terms and assume a non-trivial $\Delta t$ when numerically simulating our model below.  

\begin{figure}[!h]
    \centering
    \includegraphics[width=0.5\textwidth]{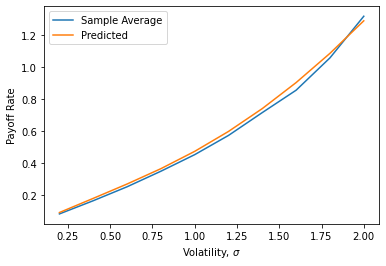}
    \caption{The predicted payoff rate (orange) with $\Delta t = 0.25, P = 1$ as compared with the empirical average of 10,000 sample trajectories (blue). Linearity is observed within a reasonable range of volatility, with nonlinear effects becoming more prominent at $\sigma > 1$.}
    \label{fig:payoff_vol}
\end{figure}

\paragraph{Temporal Evolution of Payoff Rate.} The analysis thus far concerns payoff rate at a single slice of time in which an initial price is assumed. To examine how payoff rate varies over time, we must consider the ensemble of price trajectories. This endeavor is made simple by the fact that price movement is stationary and Markovian (according to our model), so the ensemble average of the payoff rate at time $t$ is nothing but the average of the rate at all prices, weighted by the probability that price $P_t$ is reached at that time:
\begin{align} 
d\Tilde{F}(t) &:= E_{P_t}[dF(P_t)] \label{eq:payoff_rate_time}
\end{align}

Let $p(P, t)$ be the time-dependent probability density function of prices. Since $P_t$ is described by the stochastic differential equation \cref{eq:geom}, $p(P, t)$ can be obtained with the corresponding Fokker-Planck equation: 
\begin{align}
    \partial_tp(P, t) &= - \partial_{P}[\mu(P,t)p(P, t)] +  \frac{1}{2}\partial_{P}^2[\sigma^2(P,t)p(P, t)] \label{eq:fokker_planck}
\end{align}
The analytical solution to \cref{eq:fokker_planck} is obtained in \cite{stojkoski_generalised_2020} by using the Laplace-Mellin transform method. We present here the particular solution where $\mu = \sigma^2/2$ and $p(P, 0) = \delta_{P_0}$, where $\delta_{P_0}(P_0) = 1$ and $0$ elsewhere:
\begin{align}
    p(P,t) = \frac{1}{P\sigma \sqrt{2 \pi  t} } \exp \left(- \frac{ \log^2 \frac{P}{P_0}}{2\sigma^2 t} \right)
\end{align}
We can now solve for \cref{eq:payoff_rate_time}, by integrating over $P$. Again, we scale by one over liquidity to standardize comparison:
\begin{align}
    E[d\Tilde{F}(t)/L] &= \frac{\gamma}{\sigma \sqrt{2 \pi  t} } \int_0^{\infty} P^{-5/2} \left[ 2A\phi \left( \frac{B}{A} \right) + 2B\Phi \left(\frac{B}{A} \right) - B \right]  \exp \left(- \frac{ \log^2 \frac{P}{P_0}}{2\sigma^2 t} \right) dP
    \label{eq:payoff_rate_time2}
\end{align}
To calculate the total payoff over time, we integrate over \cref{eq:payoff_rate_time2}:
\begin{align}
    E[\Tilde{F}(T)/L] &= \int_{0}^{T} E[d\Tilde{F}(\tau)/L] d\tau \label{eq:ft}
\end{align}
The final formulation is unruly and thus is not included. Note that when volatility is within a manageable range, the nonlinear terms vanish, allowing \cref{eq:ft} to simplify to just $\gamma \beta T$.  

\begin{figure}[!h]
    \centering
    \includegraphics[width=0.5\textwidth]{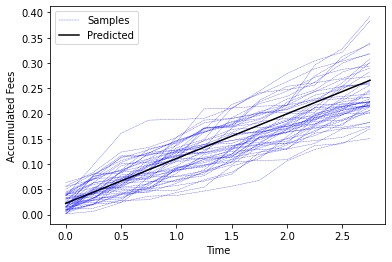}
    \caption{The expected payoff trajectory (black) from integrating over \cref{eq:ft} with $\Delta t = 0.25, T=3.0, P_0 = 1$, incorporating price diffusion through adaptive-rejection sampling of $p(P,t)$. In expectation the total payoff remains relatively linear over time, although each stochastic realization (blue) may vary greatly around the mean due to multiplicative noise added to price. }
    \label{fig:expected_traj}
\end{figure}

Despite the noisiness of realized trajectories, the stationarity of the underlying price movements implies that total payoff, being a sum of stationary mapping (ref. \cref{eq:payoff}) of prices over time, is governed by the Central Limit Theorem. Hence, reasonable bounds can be derived for the variance of total payoff. 

\section{Discussions}

There is strong demand for volatility derivatives, whose payoff is a measure of the volatility of an underlying asset. Such instruments are useful as hedge when market sentiment is uncertain, as they help manage vega exposure in a portfolio. For instance, volatility swap is an over-the-counter contract whereby the long position holder is entitled to a payoff linear to realized volatility over a period of time, in exchange for paying a pre-determined premium to the seller at maturity. Since 2003, the Chicago Board Options Exchange (CBOE) introduced an exchanged-traded derivative of its volatility index VIX tracking the volatility of S\&P 500, which has since blossomed into the second most liquid market on the exchange \cite{peter_carr_volatility_nodate}.

The volatility derivative market is still nascent in the crypto market, due in part to the difficulty of implementation and the lack of liquidity. We discussed above the intriguing discovery that fees from liquidity provision, a well-adopted and familiar practice in the crypto space, is in theory almost linear to the underlying volatility. Moreover, since the payoff rate responds to instantaneous volatility changes, the provider can benefit from dynamic hedging of risk. Hence, there is significant market potential in isolating the fees cash flow and packaging it into a volatility product.

To this end, we propose the creation of a derivative market consisting of two parties: buyers and sellers of fees cash flow, and refer to the new derivative as  \emph{liquidity fees swap}. Sellers of liquidity fees swap lock away an agreed amount of the underlying tokens in a designated automated market maker for a fixed duration. Instead of collecting the liquidity fees for themselves, they instead sell it to buyers, who pay an upfront premium in exchange for rights to future cash flow. 

The seller thus takes on a straddle-like position where any price drift negatively affects their final payoff, whereas the buyers gains a positive cash flow with near-constant vega and near-zero delta. Unlike conventional volatility products between two parties, the payoffs from liquidity fees swap is not net-zero, as fees will be extracted from noise traders through liquidity provision. Excess volatility from uninformed and non-directional traders contributes directly to the payoff of swap buyers, without affecting the payoff of the sellers. 

Additionally, we speculate that such a market will have a stabilizing effect on the primary market between the underlying assets. Any marginal increase in volatility will increase the demand for liquidity fees swap, which in turn raises the premium offered to sellers, drawing in liquidity to buffer against further price movements. If price drift were to manifest from the increased volatility (e.g. in the event that a future announcement has either positive or negative effects on price), sellers with accurate private information regarding the future relative price can profit from the excess premium provided that they hedged their impermanent loss appropriately. All in all, we believe that the introduction of this crypto-native volatility product will help accelerate the maturation of primary and derivative crypto markets.    


\printbibliography

\end{document}